# Ion-Mediated Structural Discontinuities in Phospholipid Vesicles

Judith U. De Mel[1*], Sudipta Gupta[1*], and Gerald J. Schneider[1,2*]

[1]*Department of Chemistry, Louisiana State University, Baton Rouge, LA 70803, USA*

[2]*Department of Physics & Astronomy, Louisiana State University, Baton Rouge, LA 70803, USA*

## Abstract

Despite intense research, methods for controlling soft matter's spontaneous self-assembly into well-defined layers remain a significant challenge. We observed ion-induced structural discontinuities of phospholipid vesicles that can be exploited for controlled self-assembly of soft materials. We used dynamic light scattering, zeta-potential measurement, cryo-electron microscopy, small-angle X-ray, and small-angle neutron scattering. All the experimental observations indicate that declining solvent quality and increasing osmotic pressure direct lipids to expel preferentially to the inner compartment. Upon reaching a critical concentration, excess lipids can form a new bilayer. This spontaneous self-assembly process causes simultaneous shrinkage of the aqueous core and expansion of the vesicle. This approach opens an intriguing path for controlling the self-assembly of bioinspired colloids, which can also serve as a vehicle to control the polymerization of multilayer polymeric systems.

## Introduction

As much as 70% of the earth's surface is covered by water, but 97% of all the water on earth is saline. [1] Oceanic salt concentrations range from 400 to 450 mM, whereas freshwater sources such as rivers and lakes contain minimal salts of 0.25 mM. [2, 3] In hypersaline environments such as the Dead Sea, which is about ten times saltier than the ocean, sodium and chloride collectively amount to 78% of the salinity. [4] Concerning the effect of salt on living cells, bacteria such as *E.coli* have a cytosolic NaCl concentration of about 5 mM, while the human cytosolic concentration ranges from 4 to 12 mM. [5-7] Human blood contains salt in the 110-150 mM range. [7] The tolerance of living cells to salinity differs significantly based on factors such as homeostasis mechanisms and environmental adaptations. For example, high serum sodium in humans is related to complications such as hypernatremia, cardiovascular diseases, and failures in the nervous system. [8-11] Low serum sodium is associated with hyponatremia. [10] Therefore, it is vital to understand the effects of extracellular salt on phospholipid membranes, particularly at the outer lipid membrane, which occupies the largest surface area of the living cells.

In this context, we focus on sodium chloride (NaCl), which is one of the most abundant salts that can be found in both biotic and abiotic environments, such as extracellular, intercellular fluids of living organisms and the earth's oceans. Concerning biological functioning, inter-bilayer forces are known to be modified in the presence of monovalent salts, which affect different biological processes, like cell fusion and secretion. [12] However, the exact nature of interactions with unilamellar membranes and the overall effect on the vesicle structure is still unclear. The dipolar nature of the head groups and the dielectric gradient across the membrane interphase will contribute to ion-lipid interactions in the presence of salt. Hereafter, we concentrate on zwitterionic or "neutral" lipids with phosphatidylcholine (PC) headgroups – the most abundant phospholipids in mammalian cells. [13]

Electrically neutral PC-based membranes are known to attract one another by weak van der Waals forces due to charge fluctuations. [14, 15] The repulsion between the lipid bilayers can originate from the thermal undulation of the membrane, the electrostatic interaction between the charged groups, and the hydration energy of the polar head groups manifested as the hydration pressure. [15-18] For neutral molecules, the thermal undulation and hydration energy will contribute to the repulsive force. The balance between attractive and repulsive forces determines the formation of stable single-layer or multilayer liposomes with an equilibrium lamellar spacing, $d$, and ultimately determines the formation of unilamellar or multilayer versicles. [17, 19] Therefore, a change in $d$ indicates a shift in the balance between the attractive and repulsive force. The presence of salt can modify both the attractive and repulsive forces at low salt content; however, the electrostatic interaction should be screened at sufficiently high salt concentration.

Here, we present experimental observations of previously unknown discontinuities in the structural parameters, like liposome diameter and membrane thickness. We explain that such concentration-dependent effects arise from structural transitions due to the formation of additional bilayers caused by increasing the osmotic pressure and reducing the solvent quality.

## Methods

### Materials

Highly purified (>99%) 1,2-di-(octadecenoyl)-*sn*-glycero-3-phosphocholine (DOPC) was purchased from Avanti Polar Lipids (Alabaster, AL, USA) and used without further purification. Biotechnology grade Sodium Chloride (NaCl) (99.9% purity) was obtained from VWR Life Sciences (Solon, OH, USA), organic solvents (HPLC grade), and $D_2O$ were purchased from Sigma Aldrich (St. Louis, MO, USA).

### Sample preparation

DOPC liposomes were prepared by dissolving DOPC lipid powder in chloroform, removing the chloroform using a rotary evaporator, and drying under a vacuum overnight. The dried lipid was hydrated using $D_2O$, and the resultant solution was subjected to freeze-thaw cycling by alternatingly immersing the flask in water at around 50 °C and placing it in a freezer at -20 °C in ten-minute intervals. Finally, the solution was extruded using a mini extruder (Avanti Polar Lipids, Alabaster, AL, USA) through a polycarbonate membrane with a pore diameter of 100 nm (33 passes) to obtain unilamellar liposomes. Liposome solutions prepared in $D_2O$ were mixed with NaCl solutions in a 1:1 ratio to achieve the desired extravehicular NaCl concentration, and a 24-hour waiting time ensured that samples were in their best equilibrium states. Unless stated otherwise, all experiments were conducted at ambient temperature (23 °C) and DOPC concentration of 0.25 wt%.

### Dynamic light scattering (DLS)

Dynamic light scattering (DLS) measurements were performed using a Malvern Zetasizer Nano ZS equipped with a He-Ne laser of wavelength, $\lambda = 633$ nm at 30 mW laser power, at a scattering angle $\theta = 173°$. The hydrodynamic radius, $R_h$, of the liposomes in each NaCl concentration was calculated using the Stokes-Einstein equation, $R_h = k_B T / (6\pi \eta_0 D)$, with the Boltzmann constant, $k_B$, the temperature, $T$, the viscosity of the solvent ($D_2O$ or NaCl solution), $\eta_0$. Four separate DLS measurements for each mixture were averaged.

### Zeta Potential

Zeta potential measurements were done using the Next Generation Electrophoretic Light Scattering (NG-ELS) instrument with extended capabilities to allow measurements at high salt concentrations. Approximately 0.5 ml of each sample was placed in a disposable 4 mm path-length cuvette with blackened Platinum electrodes, significantly reducing electrode polarization. Five measurements of each sample were carried out for 60 seconds at 8 or 10 V at 64 or 128 Hz.

### Cryo-Transmission Electron Microscopy (Cryo-TEM)

Cryogenic transmission electron microscopy (cryo-TEM) images were recorded on a Tecnai G2 F30 operated at 150 kV. A volume of ten microliters of the sample (0.125 wt% DOPC: in pure $D_2O$, or NaCl) was applied to a 200-mesh lacey carbon grid mounted on the FEI Vitrobot plunging station, and excess liquid was blotted for 2 s by the filter paper attached to the arms of the plunging device. The carbon grids with the attached thin film of liposome suspensions were plunged into liquid ethane and transferred to a single-tilt cryo-specimen holder for imaging. By quickly plunging into liquid ethane, the vesicles are preserved at their hydrated state at room temperature. Cryo-TEM images were obtained in the bright field setting.

## Small-angle X-ray Scattering (SAXS)

Small-angle X-ray scattering (SANS) experiments were conducted at the LIX beamline at the National Synchrotron Light Source II, Brookhaven National Laboratory, and at the Bio-SAXS beamline at the Stanford Linear Accelerator Center (SLAC) facility. The samples were measured in a flow cell with an acquisition time of 1 s at the synchrotron instrument. In contrast, the samples were loaded in 1 mm borosilicate glass capillary cylinders for the lab X-ray with an acquisition time of 10 s. The recorded intensities were corrected for dark current, empty cell, and solvent (buffer) using standard procedures. [20, 21] The scattering intensity was normalized to absolute units (cm$^{-1}$) using water as the calibration standard. [22] The data modeling is explained in the supplementary information.

## Small-Angle Neutron Scattering (SANS)

Small-angle neutron scattering (SANS) experiments were conducted at the NG 7 SANS instrument of the NIST Center for Neutron Research (NCNR) at the National Institute of Standards and Technology (NIST). [23] The sample-to-detector distances, $d$, were fixed to 1, 4, and 13 m, at neutron wavelength, $\lambda = 6$ Å. Another configuration with lenses at $d = 15.3$ m, and $\lambda = 8$ Å was used to access low $Q$'s. [24] This combination covers a $Q$ - range from $\approx 0.001$ to $\approx 0.6$ Å$^{-1}$, where $Q = 4\pi \sin(\theta/2)/\lambda$, with the scattering angle, $\theta$. A wavelength resolution of, $\Delta\lambda/\lambda = 14\%$, was used. All data reduction into intensity, $I(Q)$, vs. momentum transfer, $Q = |\vec{Q}|$, was carried out following the standard procedures implemented in the NCNR macros for the Igor software package. [25] The intensity values were scaled into absolute units (cm$^{-1}$) using a direct beam. The solvents and the empty cell were measured separately as backgrounds. The data modeling is presented in the supplementary information (SI).

## Results and Discussion

Hereafter, we report on changes in the structural parameters observed after adding salt to liposomes in an aqueous solution. Since the NaCl has been added after the self-assembly of liposomes, the inner compartment is free of salt, at least at the beginning. We test how the vesicular system responds to this initial imbalance.

## Vesicle size and surface charge

Increasing the ion concentration of solutions of liposomes and water leads to higher osmotic pressures, and a continuous reduction of the vesicle diameter is the logical consequence. However, Figure 1a illustrates that at least four different regions can be distinguished. **(1)** An initial sharp size reduction with a power-law, $\phi_M^{-0.04 \pm 0.003}$, as observed from independent DLS and SANS experiments. **(2)** There is an abrupt increase in size at $\phi_{M1}$ = 8 mM but it continues to decay with same the power-law, $\phi_M^{-0.04 \pm 0.003}$. **(3)** At higher concentrations from 75 to 500 mM, we observe a constant size within experimental accuracy. **(4)** At concentrations above, $\phi_{M3}$, we find a slight increase in size with a weak power-law dependence, $\phi_M^{0.030 \pm 0.001}$.

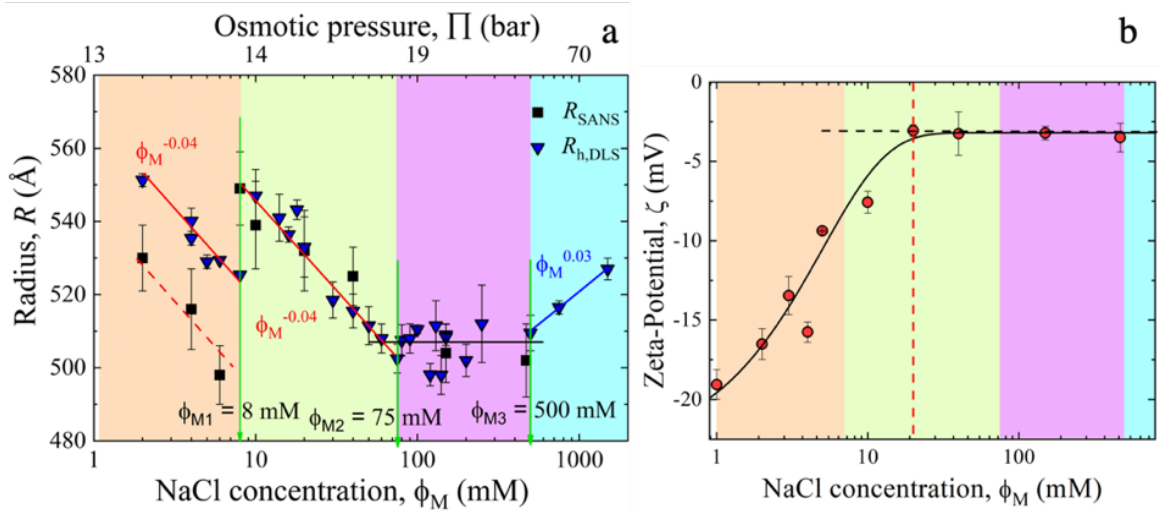

**Figure 1 – (a)** Vesicle radius in lin-log representation from dynamic light scattering (DLS) with increasing NaCl concentration, $\phi_M$. The vertical arrows indicate the transition concentrations at 8, 75, and 500 mM. **(b)** The zeta-potential data as a function of $\phi_M$. The solid line represents continuous growth. In panels (a) and (b), the four different regions are distinguished by different colors.

Figure SM4 (SI) shows a continuous linear increase of the osmotic pressure with the NaCl concentration. The continuous linear increase of the osmotic pressure, $\Pi$, with increasing the ion concentration is expected from the well-known van't Hoff law $\Pi = \varphi c R T$. [26] Here, $\varphi$ is the osmotic coefficient, $c$ is the total concentration of the Na$^+$ and Cl$^-$ ions, $R$ is the ideal gas constant and $T$ is the temperature. Hence, the observed dependence of the osmotic pressure on the ion concentration is unlikely to be the single source of the observed discontinuities.

It is well-established to assume that salt perturbs the equilibrium between electrostatic repulsion and Van der Waals attraction between the lipids. Measuring the zeta-potential (or $\zeta$-potential) provides more information about the charge on the vesicle surface. Figure 1b indicates a continuous increase of the zeta-potential with the NaCl concentration. Assuming an exponential increase can describe the data and provides a growth constant of $6 \pm 1$ mM until it reaches its plateau at around 20 mM. This 20 mM indicates a characteristic change of the zeta-potential, but none of the three discontinuities, at 8, 75, and 500 mM, seems to be connected. Hence, the zeta-potential change also appears unlikely to be related to the observed discontinuities.

As both osmotic pressure and electrostatic interaction alone are unlikely to be the sole reason, we can ask whether the balance between the osmotic pressure and Coulomb interactions may be responsible for the effects observed at 8, 75, and 500 mM. To clarify this more in detail, we investigate the structure of the vesicle as a function of the concentration in more detail, using additional techniques, cryo-transmission electron microscopy (cryo-TEM), small-angle X-ray scattering (SAXS), and small-angle neutron scattering (SANS).

### Vesicle morphology revealed by cryo-TEM

Cryo-TEM images in Figure 2 indicate a reduction of the diameter caused by the presence of salt, which is accompanied by a broadening of the circular boundary layer. This boundary layer broadening is likely related to a transition from unilamellar to multilamellar liposomes. Noteworthy, the transformation into bilamellar vesicles occurs already at shallow salt content of, $\phi_M < 20$ mM, comparable with human cytosolic salt concentration. At higher salt content $\phi_M \geq 20$ mM, cryo-TEM images reflect a change of the liposome diameter and formation of double or triple-layered multilamellar vesicles (MLVs) along with increased polydispersity. The cryo-TEM image for For $\phi_M \geq 100$ mM shows a mixture of fused and unilamellar vesicles, indicating a change to a heterogeneous system.

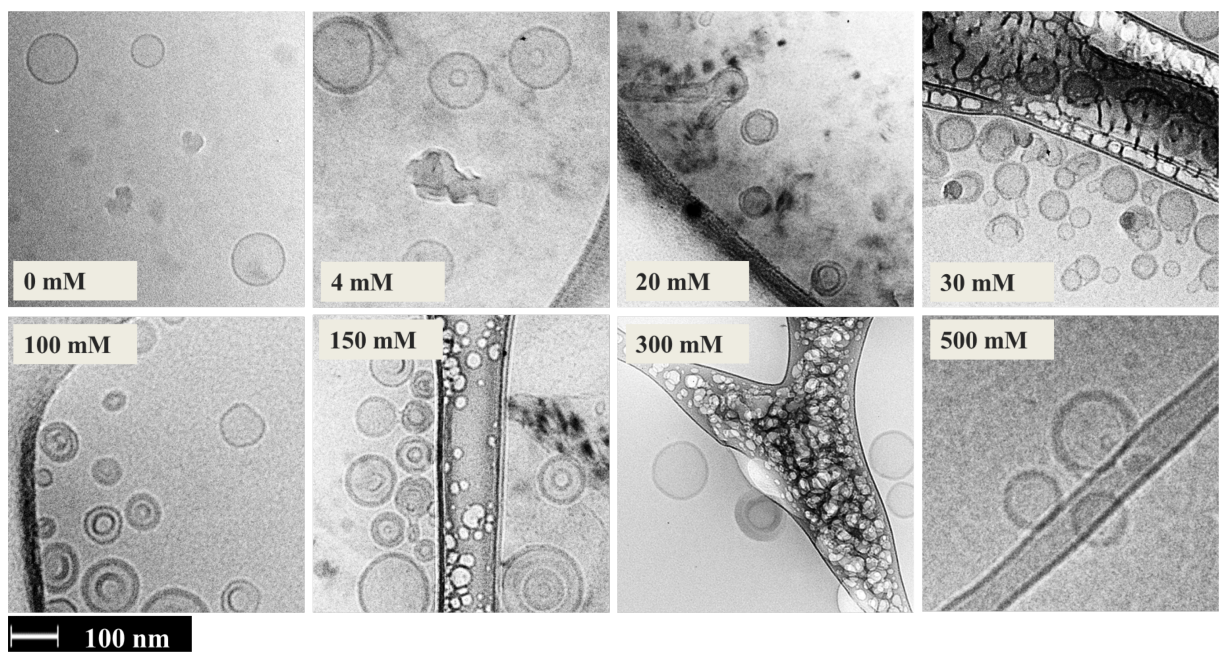

**Figure 2** – Cryo-TEM (cryogenic transmission electron microscopy) images of DOPC liposomes in D$_2$O exposed to different NaCl concentrations, as indicated in the photos. The horizontal bar on the bottom represents 100 nm.

While the cryo-TEM experiments already indicate the influence of salt on the structure, including the diameter and the number of bilayers, scattering techniques can add more information on the statistical relevance. Hence, conducting scattering experiments is relevant to show whether enough liposomes show the transition observed by cryo-TEM to be connected to the discontinuities illustrated by Figure 1 (a). Hereafter, we present both small-angle neutron scattering (SANS) and small-angle X-ray scattering (SAXS). While SANS provides essential information on the size and shape of liposomes, the scattering length and associated contrast by the phosphorous head groups for X-rays makes SAXS the perfect tool to explore changes in the bilayer, including thickness and number of lamellae.

## Vesicle structure from SANS

The SANS results for samples prepared with different salt concentrations are plotted in Figure 3a. The intensity is scaled vertically for better visualization. Scattering diagrams represent the statistical average of the morphology of liposomes, including the diameter and number of lamellar layers. The intensity vs. momentum transfer, Q, plots of the different concentrations decay from higher to lower intensity with a characteristic decay that allows us to determine the structure. Here, we used a model represented by the (black) full lines. The most apparent differences involve the growing peak intensity at higher Qs, indicating a multilamellar structure's emergence. Even at concentrations as low as 4 mM, the first signs of a statistical amount of multilamellar structures are visible. This observation is notable because 4 mM is well in the region of the human cytosolic salt concentration (4 – 12 mM).

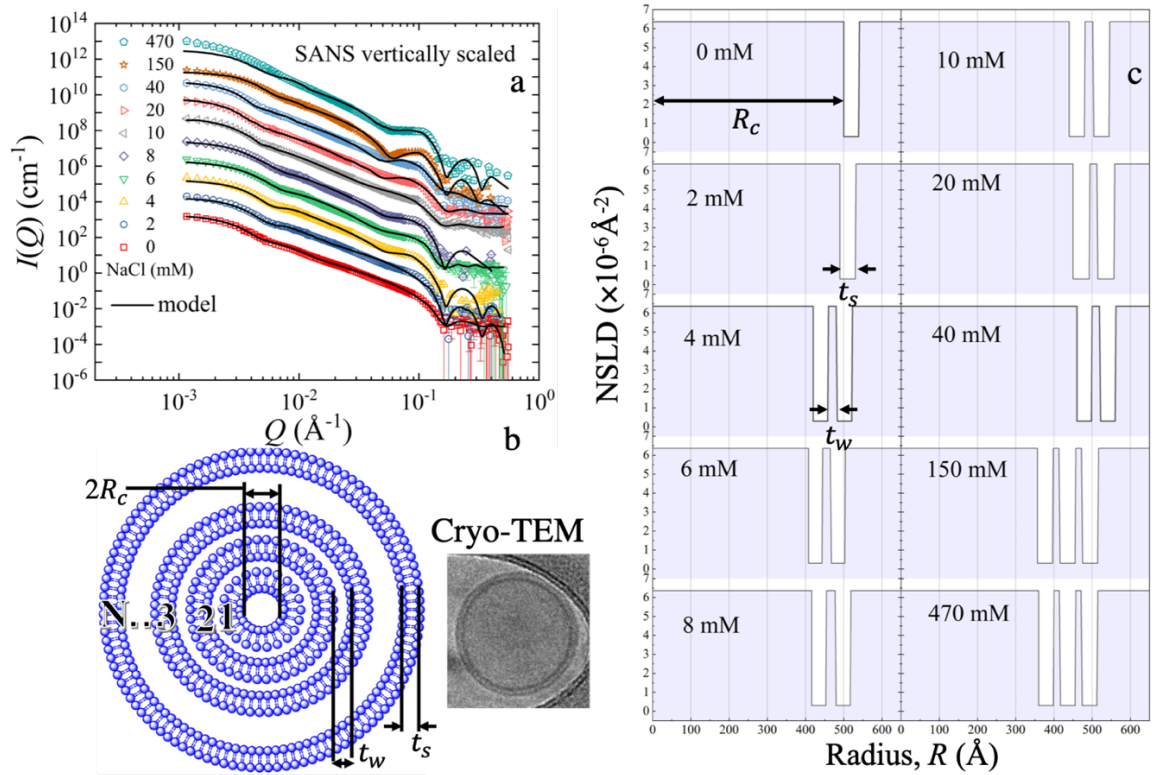

**Figure 3 – (a)** SANS scattering intensity for DOPC in $D_2O$ and different salt (NaCl) concentrations, ranging from 0 to 470 mM. The solid line represents the fits using the form factor for a unilamellar vesicle for 0 and 2 mM and a multilamellar vesicle for 4 to 470 mM samples. The data is vertically scaled for better visualization by multiplication with a constant value in a logarithmic scale. **(b)** Schematic representation of the multilamellar liposome illustrating the number of bilayers, $N$, the diameter of the core, $2R_c$, the thickness of the individual shells, $t_s$, the thickness of the interleaved solvent layers, $t_w$. Cryo-TEM image for the formation of MLVs. **(c)** Neutron scattering length density (NSLD) profile as a function of vesicle radius for different salt concentrations.

Mathematical modeling provided a more detailed analysis of the changes in vesicle radius with salt concentration. The solid lines in Figure 3a represent a multilayer vesicle form factor, with details summarized in the Data Modeling section of the SI. The model assumes a water compartment (core) of radius, $R_c$, surrounded by $N$ lipid bilayers, each of thickness, $t_s$, separated by $N$-1 inter water layers, each of thickness $t_w$.[27] The number of layers and changes in the distances from the center are illustrated by the corresponding neutron scattering length density profile in Figure 3c.

The best description of the SANS data in Figure 3a can be obtained by including a Gaussian distribution for the inter-bilayer water and lipid bilayer thickness. Numerical values from the fits are compiled in Table S1 of the SI. The width of the distribution was kept less than 0.1 for $t_s$, though a broader thickness distribution for $t_w$ ranging from 0.6 at low to 0.8 at a higher salt concentration (40 mM) is necessary. The inter-bilayer water thickness, $t_w$, is practically unchanged (~ 20 Å) for lower salt concentrations $\leq 40$ mM; however, a sharp reduction of $t_w$ to ~15 Å occurs at 150 and 470 mM.[28]

Hence, the SANS results confirm that a statistical number of liposomes changes the radius and transitions from unilamellar to multilamellar vesicles. The number of lamellae increases with the increase in concentration. However, there are three concentration regions, with 1, 2, and 3 bilayers. Before we connect multilayer formation to the size changes observed by DLS in Figure 1a, the changes in the lamellar layers are studied in more detail using SAXS.

## Membrane structure from small-angle X-ray scattering (SAXS)

While SANS provides information on the liposome diameter, small-angle X-ray scattering (SAXS) further refines our knowledge of the thickness and number of layers. Compared to neutrons, the X-ray contrast for the lipid heads with the phosphorous atoms is higher, and the instrumental resolution of X-rays is better. Hence, from a theoretical point of view, SAXS information on the bilayer structure should be more accurate.

SAXS results are presented in Figure 4. For a better comparison, the scattering intensity is vertically scaled. In the absence of salt, the first-order diffraction peak yields a repeat distance, $d = \frac{2\pi}{Q_0} = 63 \pm 1$ Å. This value is close to $63.1 \pm 0.3$ Å, previously reported for oriented stacks of unilamellar vesicles.[29] Similar to the SANS data in Figure 3, we observe a peak that grows with increasing salt concentration. Hence, the SAXS and SANS data appear to be compatible.

The Caille structure factor, with details presented in the modeling section of the SI, was used to find more information on the lamellar sheets sketched in Figure 4b.[30, 31] The results permit direct access to the lamellarity or the number of repetitive multilayers, $N$, the lamellar repeat distance, $d$, as well as the thickness of the lipid head and tail groups, represented by $\delta_H$ and $\delta_T$, respectively. The head-to-head bilayer thickness is given by $\delta_{HH} = 2(\delta_H + \delta_T)$. The position of the first-order Bragg peak is given by $Q_0$, $k_B$ is Boltzmann's constant, and $T$ is the absolute temperature. More details of the model can be found in the Modeling section of the SI.

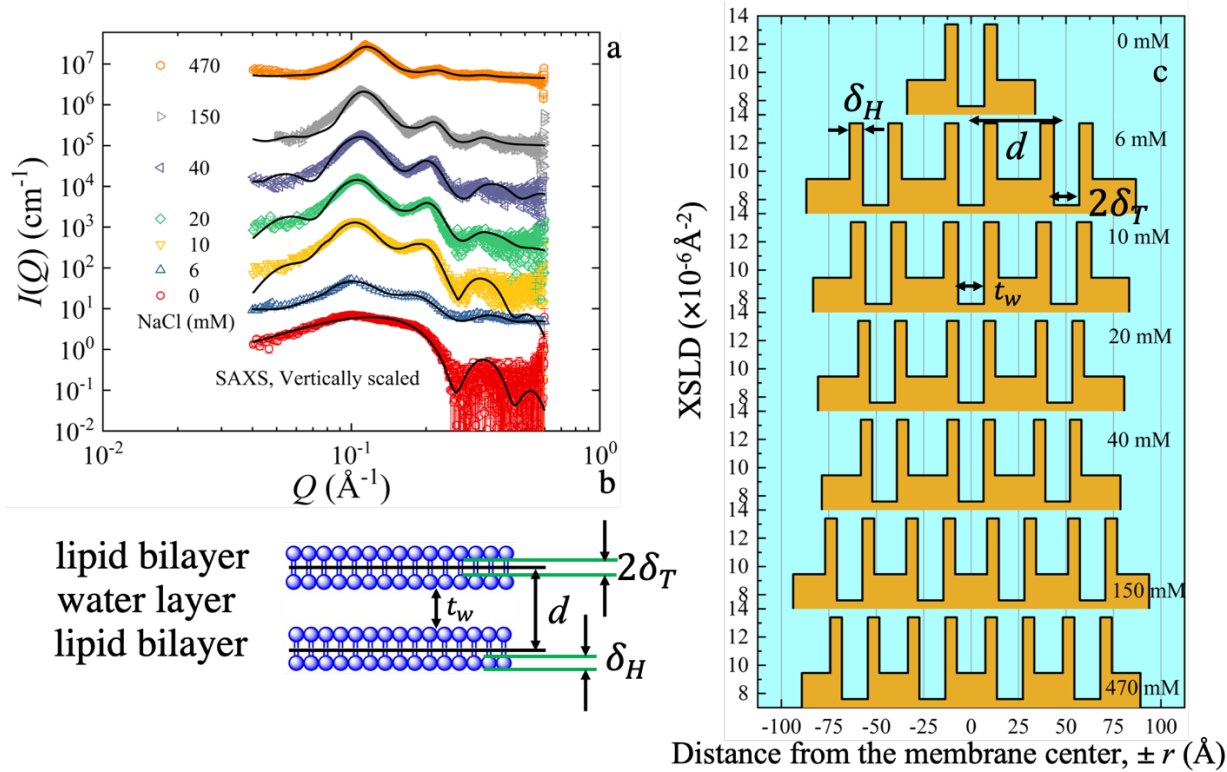

**Figure 4 – (a)** SAXS scattering intensity for DOPC in $D_2O$ with different salt (NaCl) concentrations, ranging from 0 to 470 mM. The solid line represents the fits using a lamellar structure factor. The data is vertically scaled for better visualization by multiplication with a constant value in a logarithmic scale. **(b)** Schematic representation of the lipid multilamellar structure illustrating the thickness of the lipid head, **$\delta_H$**, the thickness of the lipid tail region, **$\delta_T$**, the water layer thickness, **$t_w$**, and the lamellar repeat distance, d, of bilayers. **(c)** X-ray scattering length density (XSLD) profile as a function of distance from the membrane center for different salt concentrations.

Figure 4c shows the results of the numerical fits for different concentrations. Numerical values from the fits are compiled in Table S2 of the SI. The increase in the number of layers and changes in the distances to liposome centers with the ion concentration is illustrated by the X-ray scattering length density profile shown in Figure 4c. Additionally, we calculated the ratio of the first, second, and third peak positions, $Q_1$:$Q_2$:$Q_3$ = 1:2:3 (Figure 4a). This calculated ratio independently confirms lamellarity with well-defined repetitive distances.[32] Formation of higher-ordered lamellar structures (larger $N$) is further confirmed by a much sharper first-order diffraction peak, corresponding to more regular lamellar spacing than at low salt concentrations. The best model description of the data was accomplished with $N = 3 \pm 1$ layers, for the lower NaCl concentrations, and $N = 4 \pm 1$ layers for salt concentrations of 150 and 470 mM, respectively. The better contrast and higher resolution of SAXS enabled more layers to be resolved than in the SANS, $N = 2$ (up to 40 mM), and $N = 3 \pm 1$ (150 and 470 mM). The values agree with the statistical accuracy.

In the next step, the results are analyzed to determine whether these structural changes can explain the size decrease with the increasing salt concentration and the discontinuities at specific concentrations. Modeling parameters from SANS and SAXS are compiled in Figure 5a. While a virtually constant $\delta_{HH}$ was observed, the thickness of the water layer, $t_w$, initially decreased from 22 to 14 Å but stayed virtually constant at high concentrations (> 250 mM). The one-order of magnitude more substantial decrease of about 150 Å of the core size contributes more strongly to the entire liposome diameter, as visualized by Figure 5b. As core and liposome diameters decreased with increasing salt concentration, the discontinuities have a different origin.

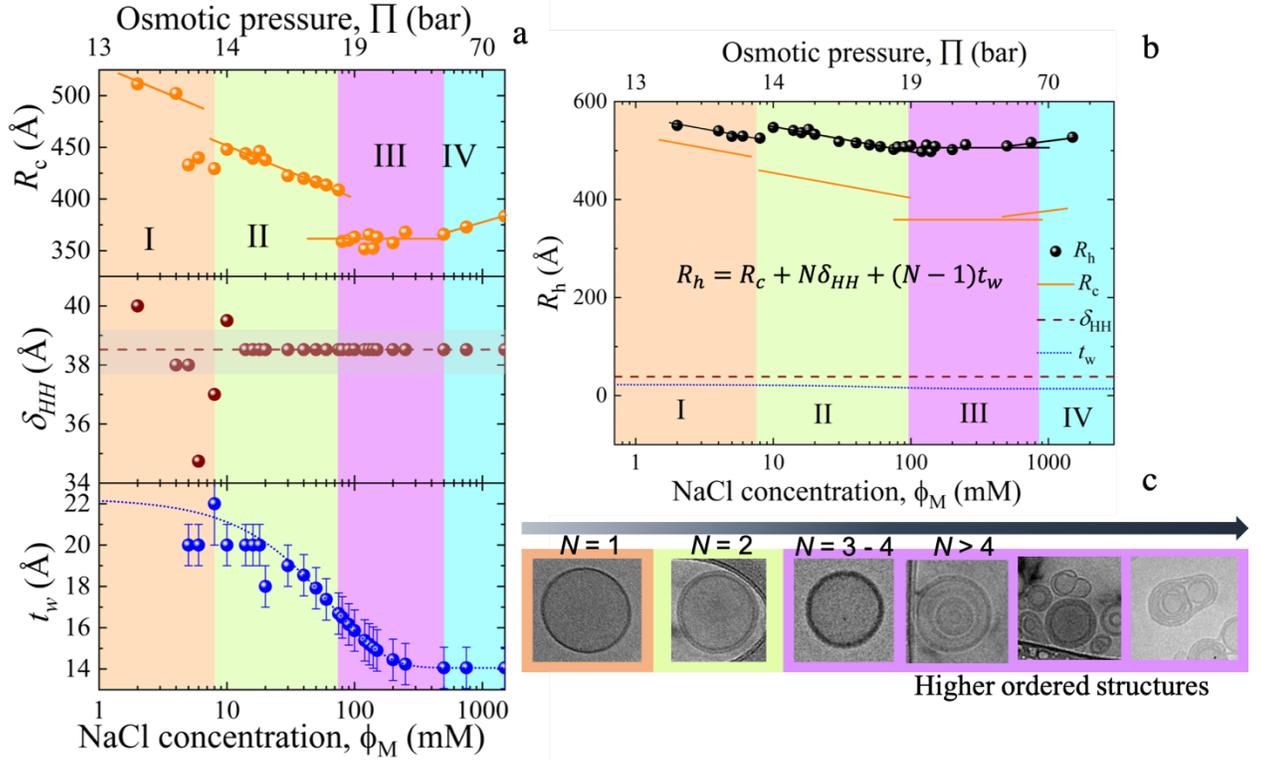

**Figure 5** – The different color codes illustrate four distinct phases with increased salt concentration. **(a)** The core radius of the vesicle, $R_c$, the bilayer thickness, $\delta_{HH} = 2(\delta_H + \delta_T)$, and the distance between the bilayers as given by the water layer thickness, $t_w$, are plotted as a function of the salt concentration, $\phi_M$, (bottom axis) and the osmotic pressure, $\Pi$ (top axis). $R_c$ follows the same power-law dependence as $R_h$ in Figure 1, represented by the solid lines. The average $\delta_{HH}$ is given by the horizontal line and its standard deviation is indicated by the shaded area. The $t_w$ exhibits a logarithmic concentration dependence that can be described by a theoretical model that relates osmotic pressure and $t_w$ by, $\Pi = P_0 \exp(-t_w/\lambda)$, for an applied pressure, $P_0$, over a decay length. [16] **(b)** The $R_h$ from Figure 1 is re-plotted to compare it with $R_h = R_c + N\delta_{HH} + (N-1)t_w$. (c) Cryo-TEM images show the evolution from ULVs to MLVs and different higher-order structures with increased salt concentration (arrow).

As illustrated by the black line in Figure 5b, the hydrodynamic radius of the liposome was calculated from $R_h = R_c + N\delta_{HH} + (N-1)t_w$. This expresses the importance of the formation of multilayers for the size discontinuities. Independent observations of the formation of these multilayers by cryo-TEM are illustrated by the images in Figure 5c. The formation of multilayers explains the apparently contradicting observation of the simultaneous expanding liposome and shrinking core diameter at 8 mM, which is also the reason for the diameter decrease with increasing concentration and the sudden increase at 8 mM.

The water layer thickness, $t_w$, decreased with increasing salt concentration. A theoretical description can be based on the observations for zwitterionic lipid bilayers that below the equilibrium bilayer separation, the inter-bilayer repulsive force falls of exponentially over a decay length, $\lambda$, which leads to $\Pi = P_0 \exp(-t_w/\lambda)$, with net repulsive pressure, $P_0$.[16, 33, 34] The corresponding pressure distance plot that shows fitted data is presented in Figure SM 5, SI. A decay length, $\lambda$, of around 1 Å was determined by fitting the data. Inter-bilayer water thickness values in the 14 to 20 Å range correspond to $P_0 = (4.9 \pm 1.0) \times 10^7$ bar. From this value, $\Pi = 40$ bar at equilibrium has been calculated using an inter-bilayer distance of 14 Å, at a salt concentration of 470 mM. The osmotic pressure is 39 times higher than the normal atmospheric pressure. Therefore, at high salt concentrations, Figure 5b reflects the diverging nature of the repulsive force, preventing individual bilayers from coming in close contact. To maintain a balance with the osmotic force, the inter-bilayer repulsive force increases exponentially, $\exp(-t_w/\lambda)$, accompanied by the formation of higher-order MLVs. The influence of the osmotic pressure on the bilayer thickness, $\delta_{HH}$, is negligible, cf. Figure SM4, SI. Thus, the lamellar repeat distance, $d = t_w + \delta_{HH}$, follows the concentration dependence of $t_w$. The influence of $t_w$ on the liposome diameter, $d$, is less important, because of $t_w \ll R_c$.

Hence, the experiments provide a plausible explanation for the observed size reduction and discontinuities in DOPC with increasing the salt concentration.

The increase of the liposome diameter and the decrease of the core size is well compatible with the emergence of new bilayers, with the remarkable result that individual layer thickness stays virtually constant. The thickness of the inter-bilayer water shows a continuous concentration dependence instead of abrupt changes observed in the liposome radius. Hence, we conclude that the observed continuous decrease of the diameter by DLS is a consequence of the change of the inter-layer, while the formation of new bilayers causes discontinuities.

With this correlation between diameter and molecular structural parameters, the molecular parameters are explored in more detail in the next step. The first observation of a continuous change in the thickness of the inter-bilayer watermarks is already a fundamental key observation, which requires a pressure difference between inter-bilayer and bulk water. As the experiments showed the formation of MLVs, ions could intrude into the inter-bilayer water during this re-assembly. However, the thickness decrease requires a positive pressure, hence a lower salt concentration inside than outside.

To answer the question of ions in the inter-bilayer water in more detail, the dependence of the diameter decreases on the concentration below and above 8 mM, which can be compared. The experimentally observed exponents of the power laws that describe the diameter reduction of the core below and above 8 mM are the same within the experimental accuracy. Initially, there is no salt inside the membrane, and there is no reason that salt intrudes into the intact membrane below 8 mM. Since the exponent is the same for the ranges < 8 mM and 8 mM < $\phi_M$ < 75 mM, the repulsive force seems unchanged in region II. Hence, the observation of the same exponent implies that the concentration difference below and above 8 mM is the same, which leads to a concentration of salt equal to zero in the core.

Within region two, an exponent for the inter-bilayer water layer follows a power law $\Phi_M^{-0.15}$. This is one order of magnitude steeper than the core or liposome. Given the limited concentration window, we want to interpret this value sparingly. However, because this change is at least of comparable order of magnitude with the core shrinkage, the salt or salt enrichment in the inter-bilayer water layer for $\phi_M < 75$ mM can be excluded.

The constant thickness of the inter-bilayer water is a consequence of salt intrusion, which causes the cryo-TEM observation of mixtures of unilamellar and fused vesicles at $\phi_M \geq 100$ mM (regions III and IV in Figure 5). The salt concentration in the water inside and outside the vesicles is the same in regions III and IV.

The results presented in Figure 5 establish the formation of multilayers as the origin of the size discontinuities. In the next step, the individual bilayers are examined. For more straightforward wording, the numbers 1, 2, and 3 indicate the inner to the outermost bilayer. The changes in the distances of the bilayers, $R_1$, $R_2$, and $R_3$, from the center of the liposomes as a function of the salt concentration are illustrated in Figure 6a.

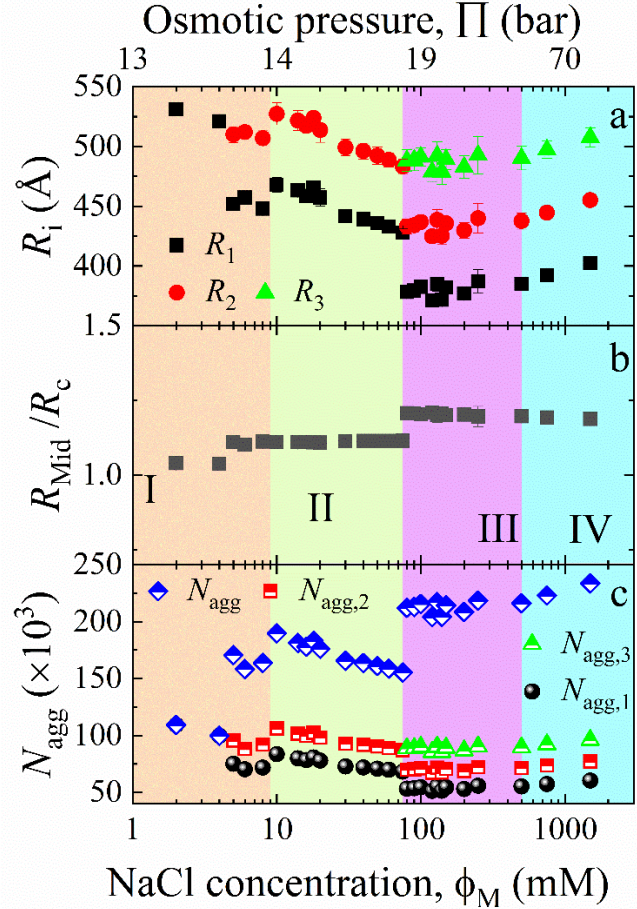

**Figure 6** – (a) The distances $R_1$, $R_2$, and $R_3$ of the center of bilayer 1, bilayer 2, and bilayer 3, respectively, from the center of the liposome. **(b)** The distance of the bilayer midplane from the vesicle center, $R_{Mid} = R_c + N \delta_{HH}/2 + (N-1) t_w/2$, as a function of salt concentration, $\phi_M$. **(c)** The aggregation number, $N_{agg}$, given by the number of lipid molecules per vesicles is presented as a function of $\phi_M$, and the osmotic pressure, $\Pi$. A comparison has been made to the number of lipids in the inner, middle, and outer shells of a MLV given by, $N_{agg,1}$, $N_{agg,2}$, and $N_{agg,3}$, respectively. Here, $N_{agg} = N_{agg,1} + N_{agg,2} + N_{agg,3}$.

Only one bilayer exists below 8 mM (region I) with the radius $R_1$. In the range 8 to 75 mM (region II), a second layer emerges. The radius of the second layer $R_2$ is greater than the radius of $R_1$ in the region I (< 8mM). However, $R_1$ in region 2 is lesser than $R_1$ in region 1. This already mirrors that the existence of layer 2 is responsible for the size discontinuity observed by DLS (Figure 1). Inspecting region III from around 75 to 500 mM reveals a change in the concentration dependence. However, continuous transitions from layer 2 to layer 3 and from layer 1 to layer 2 are observed. These results suggest that adding salt creates one layer closer to the center, but the diameter of the others stays essentially the same. Such behavior is not observed transitioning from Region III to Region IV, where $R_1$, $R_2$, and $R_3$ do not show any visible discontinuity. Instead of a size decrease a diameter increase of each bilayer is observed. Another observation is the similar concentration dependence of the different radii visible in Figure 6a.

The discontinuity is also visible in the change of the distance of the bilayer midplane from the vesicle center, $R_{Mid} = R_c + N\,\delta_{HH}/2 + (N-1)\,t_w/2$, which abruptly changes at the transition concentrations, but stays constant within the regions.

More details can be revealed by calculating the number of lipids in each bilayer, $N_{agg}$, also called the aggregation number. The numbers follow the same convention: 1, 2, 3 from the inner to the outermost layer, respectively. The total number of lipids in each liposome can be calculated as the sum of the lipids in each layer, e.g., $N_{agg} = N_{agg,1} + N_{agg,2} + N_{agg,3}$.

As illustrated in Figure 6c, numbers, $N_{agg}$, show a concentration dependence and discontinuities. The innermost layer has the minimum number of lipids, and the outermost layer has the most lipids. We also notice that the number of lipids per liposome increased with increasing the concentration, except for region II. However, the decrease in region II is less than the increase during the transition from region I to II and from II to III.

Given the observation of a size change with the concentration, the changing number of lipids could result from the geometrical packing of the lipids in the layer. Hence, Figure 7a displays the number of lipids, and Figure 7b the equivalent lipid surface density, both as a function of the distance from the liposome center.

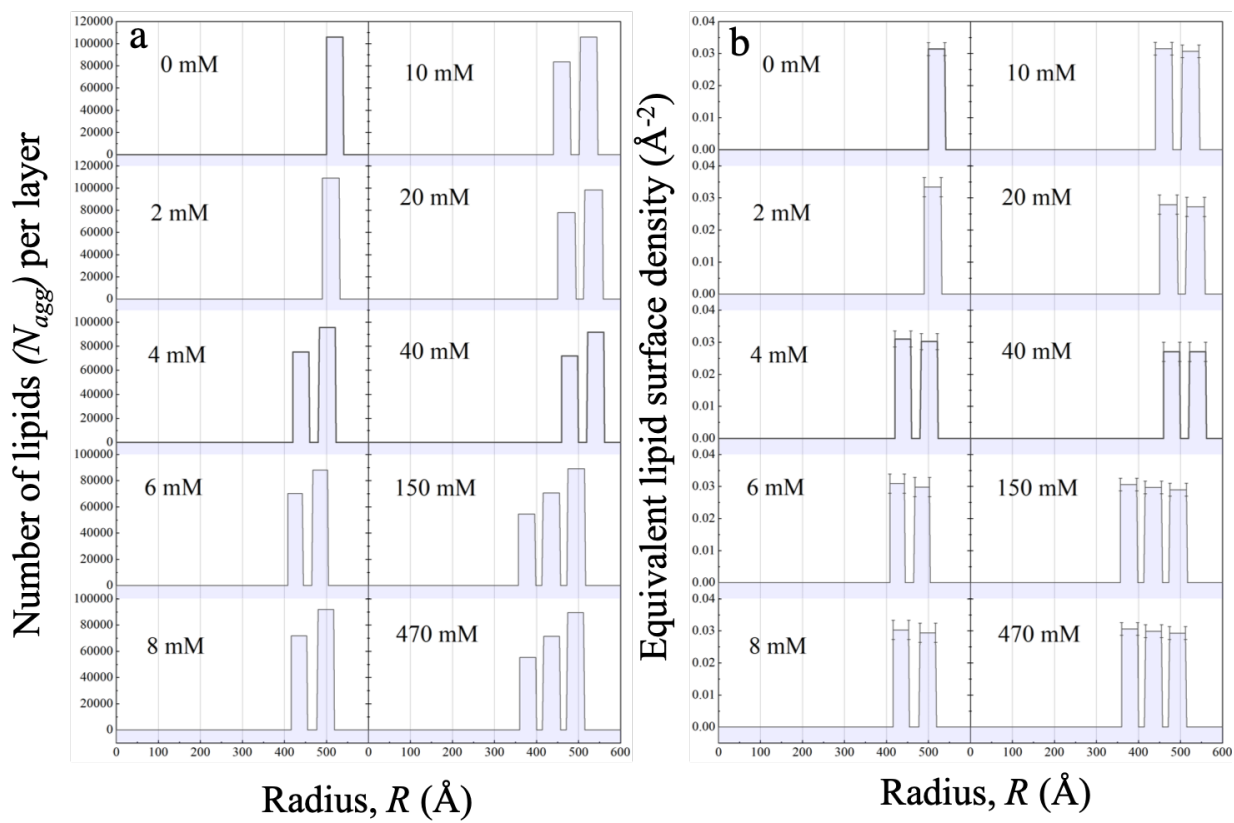

**Figure 7** – (a) The number of lipids per layer, $N_{agg}$, as a function of the radius or the distance from the center of the vesicles. (b) The equivalent lipid surface density on each layer as a function of the radius. For each layer it is calculated on the surface of the bilayer midplane and represents a uniform surface density, $0.030 \pm 0.003$ Å$^{-2}$.

While Figure 7a shows a change in the number of lipids in each layer, the equivalent lipid surface density stays constant within the experimental accuracy, $N_{agg}/(4\pi R_i^2) = 0.030 \pm 0.003$ Å$^{-2}$. Hence, it is likely that an additional layer is created to maintain the equivalent lipid surface density. Since all lipid heads occupy approximately the same space for both the inner and outer leaflets of each bilayer, the inner layer starts to relieve the pressure by releasing lipids first. [35] This explanation is also compatible with the transitions from region I to II and from region II to III lead to a layer that is continuous.

The elastic energy of liposomes is proportional to the relative change in the surface area. If the compression becomes too large, one bilayer may no longer be able to accommodate all lipids. Thus, a discontinuity may arise from a spontaneous relief by releasing lipids from the bilayer. This is supported by the fact that the lipids prefer maintaining a constant surface density in each layer. It is given by $N_{agg}/(4\pi R_i^2) = 0.030 \pm 0.003$ Å$^{-2}$. This is presented in Figure 7b. In our case, the change in the distance between the bilayers is not enough to see an influence of curvature on lipid density. The experimental data shows the emergence of new bilayers with smaller and larger diameters compared to the results for 0 mM concentration. Since the released lipids lead to an increase in the concentration of free lipids in the system, we expect these may form new bilayers. As obtained from DLS in Figure 1a, the size was reduced by 9 %, whereas the surface area of the vesicles was reduced by 18 % at 75 mM. At high salt content from 75 to 500 mM, the size and the surface area appear to be virtually constant, followed by a 3 % increase in size and a 7 % increase in surface area between 500 to 1500 mM. When the number of lipids per vesicle is approximately 2 times more than in ULVs, we observe formations of MLVs. The shrinkage in size and the formation of MLVs are intertwined. The balance between the outside osmotic pressure and the inter-layer repulsive force of the membranes can control the change in the size of the vesicles. The corresponding repulsive pressure increases dramatically to 39 atm for a salt concentration of 470 mM. However, as the SANS/SAXS results illustrate, internal layers are created even at these high concentrations.

Let's first consider one bilayer to expand this discussion toward a quantitative understanding. For DOPC it was observed that the inner and outer monolayers are identical with slight dependence on the vesicle curvature. [35] From X-ray diffraction results shown in Figure 4a, for a thickness, $\delta_{HH}$ = 41 Å, at 0 mM salt concentration, the number of lipids per vesicle in one shell for ULVs is given by, $N_{agg,1} = V_s/V_l = 1 \times 10^5$, for a vesicle of $R_h$ = 551 Å (DLS), with $V_s$ being the outer shell volume of the ULV and $V_l$ the molar volume of the phospholipid. With an increase in salt content to 150 mM the size of the ULV of thickness, $\delta_{HH} \approx 39$ Å (SAXS), shrinks to $R_h$ = 509 Å, and will have the number of lipids per vesicle, $9 \times 10^4$. Therefore, under this assumption of the formation of ULVs, there will be an excess of $2 \times 10^4$ lipid molecules per vesicle in the solution. The excess lipid concentration is equivalent to a monolayer membrane surface, that can be accommodated into MLVs. As explained below, the increasing zeta potential indicates a lower solubility of the outermost layer, further pointing to generating new layers inside the internal compartment. Figure 6c illustrates this discussion by plotting the number of lipids per vesicle aggregation number, $N_{agg}$, with salt concentration. This shows an excess number of lipids per vesicle determined by $N_{agg}$ at each transition concentration. We have $N_{agg}$ by a factor 2 higher than in the absence of salt to observe the transition to MLVs.

In the next step, we need to understand the continuous size reductions in regions I and II underlying the discontinuity at 8 mM. If we compare the effect of salt on vesicle size above and below the transition concentration $\phi_{M1} = 8$ mM in Figure 1a, we observe a power-law decay of the vesicle size for $\phi_M \leq 80$ mM is observed with the same power law exponent in the entire region.

The number of layers, $N$, in our MLVs, is almost constant in region two of the semi-dilute salt concentrations, $\phi_{M1} < \phi_M < \phi_{M2}$. In addition to the existence of weak attractive van der Waals force between the lipid layers which only dominates at very low salt concentrations, another possibility is hydration attraction or H-bonding across a water layer of thickness $t_w$ due to complementary surface polar head groups. [36] Rand et al. theoretically predicted such a mechanism in understanding the membrane's surface hydration. [36] In this case, partial dehydration of the lipid heads should further facilitate hydration attraction at high salt concentrations.

At low salt concentrations, the zeta potential data also show a net decrease of the negative charge on the vesicle surface which reaches a plateau at 20 Mm (Figure 1b). The growth constant $6 \pm 1$ mM is close to the first transition concentration, $\phi_{M1} = 8$ mM. The sign and the magnitude of the zeta potential are determined by the net charge deposited on the vesicle's surface. The negative zeta potential of DOPC vesicles at very low salt concentration results from the negatively charged phosphate groups being exposed to the outer surface. The polar head groups are known to reorient with the increasing ionic strength. [37] This phenomenon is further facilitated by the osmotic pressure. Therefore, $Na^+$ ions bind to the phosphate groups of the lipid head and $Cl^-$ ions bind to the trimethylammonium, $N^+(CH_3)_3$. This causes an increase in surface charge density at the interface of water and the polar lipid head group (Figure 1c). The increasing zeta potential might cause localized surface insolubility which further contributes to the formation of MLVs. A comparison of the vesicle size and zeta potential from Figure 1a and b, respectively clearly illustrates a decrease in vesicle size and increase in the surface charge for concentration $\lesssim 20$ mM, where MLVs have formed. The corresponding size polydispersity from DLS increases with increasing salt concentration as presented in the Figure SM2, SI.

The SAXS data reveal how the interplay between the attractive and repulsive forces in the phospholipid bilayer determine the structure. Two parameters are particularly sensitive, the lamellar repeat distance, $d$, and the bilayer thickness, $\delta_{HH}$. With an increase in NaCl content the $Cl^-$ and $Na^+$ ions can associate with the trimethylammonium and phosphate groups of the polar lipid head causing an effective decrease in the dipole potential of the PC lipid membrane. [38] This causes an increase in electrostatic repulsion between the charged surfaces that overcomes the weak van der Waals attraction. These observations are supported by $\sim 3\%$ swelling of $\delta_{HH}$, and $\sim 1\%$ decrease in $d$, respectively for salt concentrations $\leq 10$ mM. With further increase in NaCl content, the arrival of the new $Cl^-$ ions start to screen the existing electrostatic repulsion between the surfaces. This causes a sharp shrinkage of $\delta_{HH}$ by $\sim 5\%$ and reduction of $d$ by $\sim 13\%$ at 470 mM.

## Summary

The study demonstrates how a seemingly simple and often overlooked environment with enhanced salt or ion concentration can be used to transform unilamellar to multilamellar vesicles while controlling the overall size of the vesicle and water core simultaneously even at sub physiological concentrations. Our experimental results provide a plausible explanation other than the previously hypothesized balance between osmotic pressure and electrostatic interactions. One reason for the observation could be the experimental condition in which the salt was added after the self-assembly of the liposomes, i.e., only from the outside. Hence, initially ($t = 0$ s), salt can only be outside the membrane, not inside or in the inner water compartment. The experiments showed multiple transitioning stages of the neutral phospholipid vesicles with abrupt phase transitions at $\phi_{M1} = 8$ mM, $\phi_{M2} = 75$ mM, and $\phi_{M3} = 500$ mM concentrations, which can be indicators for the formation of MLVs and higher ordered hybrid structures in a saline environment window ranging from very low concentrations as in freshwater to very high as in water of the Dead Sea. Finding the continuous size change and abrupt phase transitions at specific concentrations are likely to enhance understanding of cell signaling, translocation, intracellular biological functioning, and cell division in the high saline environment but also will help design lipid drug delivery vesicles with controlled membrane transport properties with the strong dependence on the number of layers as a response to external salinity variation.

Furthermore, since the permeation of molecules through layers is determined by the thickness and the number of layers, our results show that a change in the salt concentration is expected to affect the permeation rate. This will help to manipulate existing biocompatible materials and provide a better understanding of concentration-dependent permeation rates. For example, a novel pathway for controlling the encapsulation efficiency above 43% is demonstrated, which was achieved using different phospholipids. [39]

## Supplementary information

Accompanies this paper.

## Supplementary information

Corresponding author and ORCID

Judith U. De Mel – Department of Chemistry, Louisiana State University, Baton Rouge 70803, Louisiana, United States 0000-0001-7546-1491

Sudipta Gupta – Department of Chemistry, Louisiana State University, Baton Rouge 70803, Louisiana, United States 0000-0001-6642-3776

Gerald J. Schneider − Department of Physics &Astronomy and Department of Chemistry, Louisiana State University, Baton Rouge 70803, Louisiana, United States; orcid.org/0000- 0002-5577-9328; Email: gjschneider@lsu.edu

## Competing financial interests:

The authors declare no competing financial interest.

## Acknowledgments

The U.S. Department of Energy (DOE) supports the neutron scattering work under EPSCoR Grant No. DE-SC0012432 with additional support from the Louisiana Board of Regents. The Center for High-Resolution Neutron Scattering provided access to the small-angle scattering instruments, a partnership between the National Institute of Standards and Technology and the National Science Foundation under Agreement No. DMR-1508249. We thank Lin Yang and Shirish Chodankar from 16-ID, LIX beamline at National Synchrotron Light Source (NSLS) II. The LiX beamline is part of the Life Science Biomedical Technology Research resource, primarily supported by the National Institute of Health, National Institute of General Medical Sciences under Grant P41 GM111244, and by the Department of Energy Office of Biological and Environmental Research under Grant KP1605010, with additional support from NIH Grant S10 OD012331. As a NSLS II facility resource at Brookhaven National Laboratory, work performed at Life Science and Biomedical Technology Research is supported in part by the US Department of Energy, Office of Basic Energy Sciences Program under Contract DE-SC12704. We thank Thomas Weiss from BL 4-2 at Stanford Synchrotron Radiation Lightsource for assisting with initial SAXS experiments. Use of the Stanford Synchrotron Radiation Lightsource, SLAC National Accelerator Laboratory, is supported by the U.S. Department of Energy, Office of Science, Office of Basic Energy Sciences under Contract No. DE-AC02-76SF00515. The SSRL Structural Molecular Biology Program is supported by the DOE Office of Biological and Environmental Research, and by the National Institutes of Health, National Institute of General Medical Sciences (including P41GM103393). The contents of this publication are solely the responsibility of the authors and do not necessarily represent the official views of NIGMS or NIH. We would like to thank Rafael Cueto (LSU) for his support in DLS experiments, Jibao He (Tulane University, USA) for his support in Cryo-TEM, and John Miller (Enlighten Scientific LLC, USA) for conducting zeta potential measurements. We thank Professor Jayne Garno, Chemistry Department, Louisiana State University, for carefully proofreading.

# Ion-Mediated Structural Discontinuities in

# Phospholipid Vesicles

Judith De Mel[1], Sudipta Gupta[1], and Gerald J. Schneider[1,2]

*[1]Department of Chemistry, Louisiana State University, Baton Rouge, LA 70803, USA*

*[2]Department of Physics & Astronomy, Louisiana State University, Baton Rouge, LA 70803,*

*USA*

## Supplementary Information

## Data Modeling

**Bilayer structure**: The random lamellar sheet consisting of the heads and tails of the phospholipids can be modeled using the Caille structure factor. [S1, 2] It provides direct access to the macroscopic scattering cross-section given by the scattering intensity for a random distribution of the lamellar phase as

$$\frac{d\Sigma}{d\Omega}(Q) = 2\pi \frac{VP(Q)S(Q)}{Q^2 d} \tag{S1}$$

with the scattering volume, $V$, and the distance of the lamellae, $d$. The form factor is given by:

$$P(Q) = \frac{4}{Q^2} \left[ \Delta\rho_H \left\{ \sin\left(Q(\delta_H + \delta_T)\right) - \sin(Q\delta_T) \right\} + \Delta\rho_T \sin(Q\delta_T) \right]^2 \tag{S2}$$

The scattering contrasts for the head and tail are $\Delta\rho_H$ and $\Delta\rho_T$, respectively. The corresponding thicknesses are $\delta_H$ and $\delta_T$, respectively, as presented in *Figure SM1*. The head-to-head bilayer thickness is given by, $\delta_{HH} = 2(\delta_H + \delta_T)$. The Caille structure factor is given by

$$S(Q) = 1 + 2\sum_{n=1}^{N-1}\left(1 - \frac{n}{N}\right)\cos(Qdn)\exp\left(-\frac{2Q^2 d^2 \alpha(n)}{2}\right) \tag{S3}$$

with the number of lamellar plates, $N$, and the correlation function for the lamellae, $\alpha(n)$, defined by

$$\alpha(n) = \frac{\eta_{cp}}{4\pi^2}(\ln(\pi n) + \gamma_E) \tag{S4}$$

with $\gamma_E = 0.57721$ the Euler's constant. The elastic constant for the membranes are expressed in terms of the Caille parameter, $\eta_{cp} = \frac{Q_0{}^2 k_B T}{8\pi\sqrt{\kappa_b\,\kappa_A}}$, where $\kappa_b$ and $\kappa_A$ are the bending elasticity and the compression modulus of the membranes, respectively. Here $\kappa_A$ is associated with the interactions between the membranes. The position of the first-order Bragg peak is given by $Q_0$, and $k_B$ is the Boltzmann's constant and $T$ the absolute temperature.

**Vesicle structure:** The vesicle form factor is modeled using an extension of the core-shell model used in our previous studies.[S3, 4] The core is filled with water and in case of multilamellar liposomes encapsulated by $N$ shells of lipids and $N$-1 layers of solvent as illustrated in *Figure SM1*. Each shell thickness and scattering length density is assumed to be constant for the respective shell. The 1D scattering pattern is described by:

$$P(Q, R, t, \Delta\rho) = \frac{\phi[F(Q)]^2}{V(R_N)} \tag{S5}$$

with

$$
\begin{aligned}
F(Q) = (\rho_{\text{shell}} \\
- \rho_{\text{solv}}) \sum_{i=1}^{N}\left[3V(r_i)\frac{\sin(Qr_i) - Qr_i\cos(Qr_i)}{(Qr_i)^3}\right. \\
\left. - 3V(R_i)\frac{\sin(QR_i) - Qr_i\cos(QR_i)}{(QR_i)^3}\right]
\end{aligned}
\tag{S6}
$$

For

$$
\begin{aligned}
r_i &= r_c + (i-1)(t_s + t_w) &&\text{solvent radius before shell } i \\
R_i &= r_i + t_s &&\text{shell radius for shell } i
\end{aligned}
\tag{S7}
$$

Here, $V(r)$ is volume of the sphere with radius $r$, $r_c$ is the radius of the core, $t_s$ is the thickness of the individual shells, $t_w$ is the thickness of the interleaved solvent layers, $\phi$, the corresponding lipid volume fraction. For DOPC, we used the scattering length density of the shell, $\rho_{\text{shell}} = 3.01 \times 10^9$ cm$^{-2}$ and for D$_2$O the scattering length density of the solvent, $\rho_{\text{solv}} = 6.36 \times 10^{10}$ cm$^{-2}$, respectively.[S5] The macroscopic scattering cross-section is obtained by

$$\frac{d\Sigma}{d\Omega}(Q) = \int dr P(Q, R, t, \rho_{\text{lipo}}, \rho_{\text{solv}})s(r) \tag{S8}$$

For the size polydispersity, $s(r)$, we used a Schulz distribution and a log-normal distribution. In addition, the thickness of the shell and the solvent are convoluted with a Gaussian distribution function to account for the thickness polydispersity.

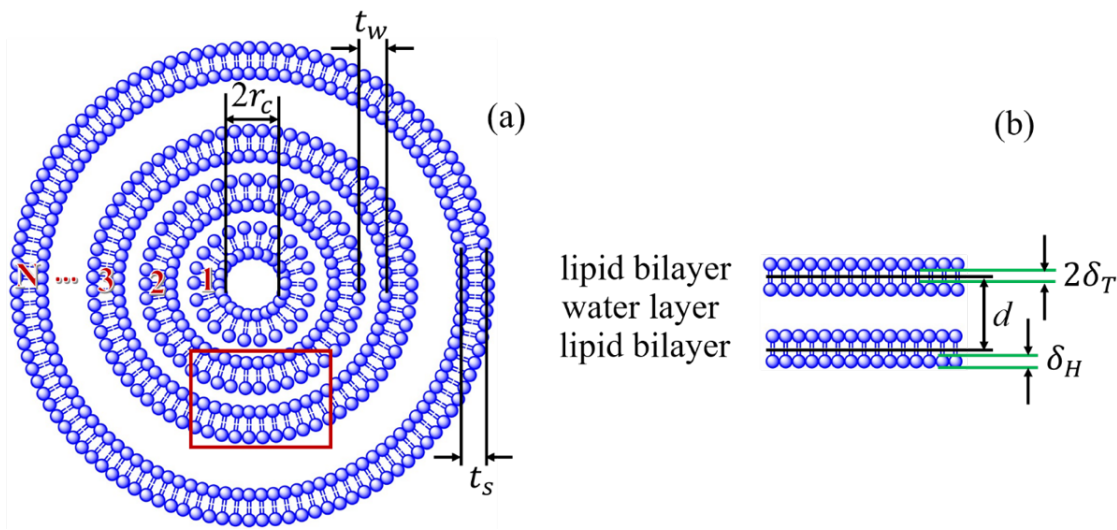

**Figure SM1** – *Schematic representation of (a) the multilamellar liposome, and (b) lipid multilayers illustrating the number of bilayers, N, the radius of the core, $r_c$, the thickness of the individual shells, $t_s$, the thickness of the interleaved solvent layers, $t_w$, the thickness of the lipid head, $\delta_H$, the thickness of the lipid tail region, $\delta_T$ and the lamellar repeat distance, d, of bilayers.*

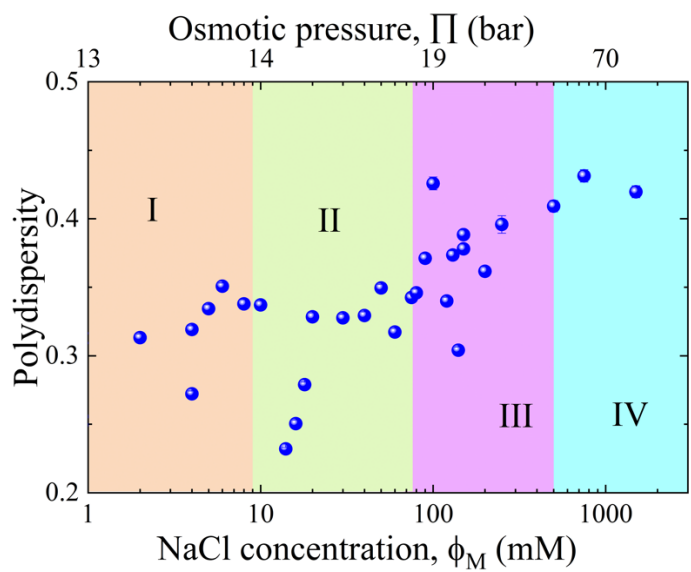

**Figure SM2** – *Illustrating the size distribution of the hydrodynamic radius from DLS as a function of the salt concentration, $\phi_M$ and osmotic pressure, $\Pi$.*

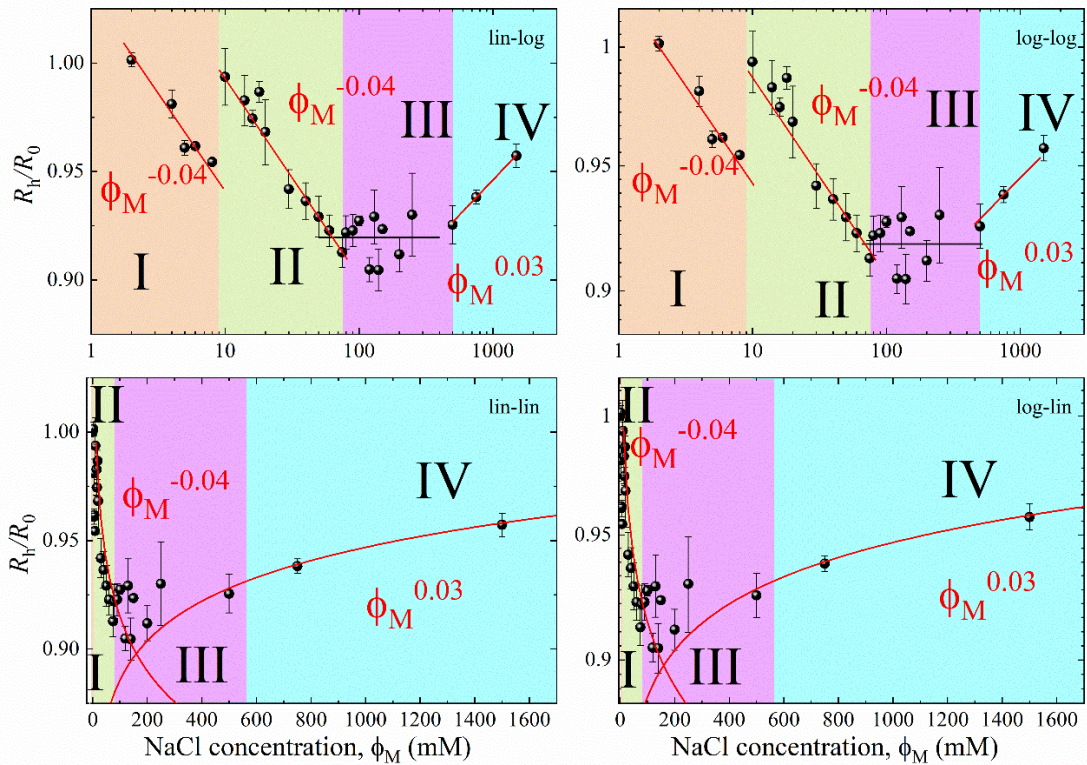

**Figure SM3** – *The normalized hydrodynamic radius, $R_h/R_0$, as a function of the salt concentration, $\phi_M$ in lin-log, log-log, lin-lin and log-lin representations. Here $R_0$ is the hydrodynamic radius without any salt. The solid lines represent the different power laws. The four different regions are marked by different colors.*

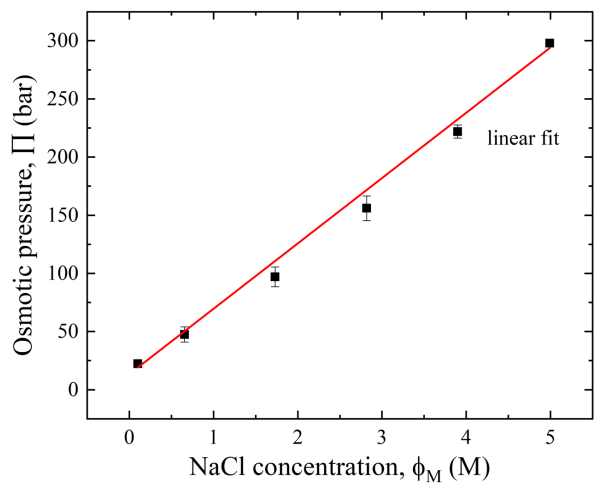

**Figure SM4** – *Linear dependence of osmotic pressure, $\Pi$ on NaCl concentration, $\phi_M$.*[S6]

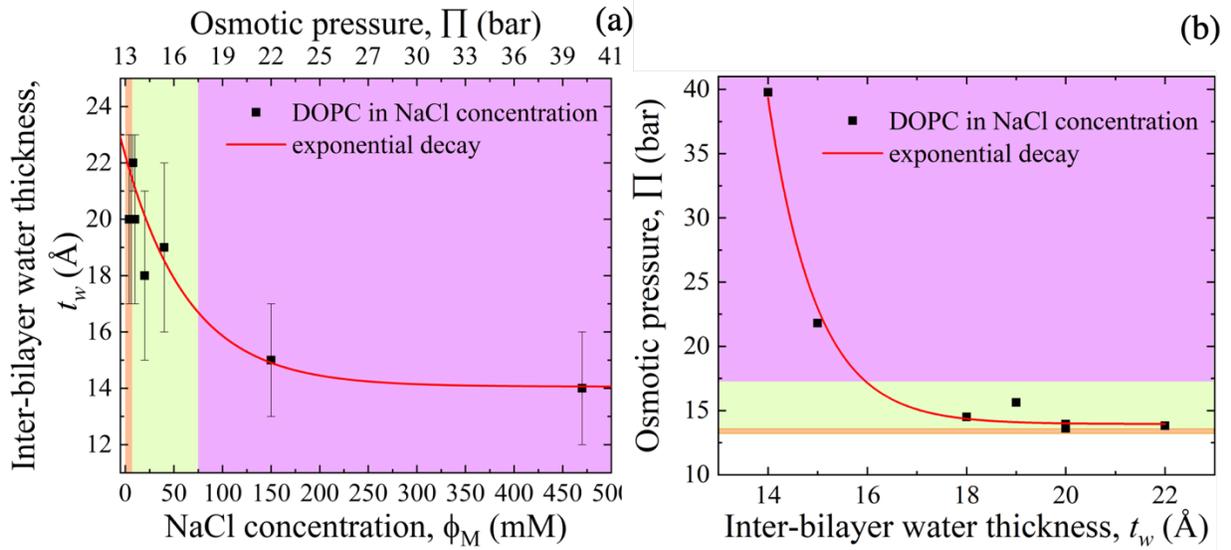

**Figure SM5** – *(a) The lamellar repeat distance, d, as a function of the salt concentration, $\phi_M$ and the osmotic pressure, $\Pi$. The solid line represents an exponential decay, the same shape as for the interlayer water layer thickness, $t_w$, illustrated in the main text. (b) Osmotic pressure, $\Pi$, as a function of the distance between the bilayers as given by the water layer thickness, $t_w$. The solid line is a fit assuming exponential decay for, $\Pi = P_0 \exp(-t_w/\lambda)$, with the magnitude of the applied pressure, $P_0$, over a decay length, $\lambda$.*

**Table S1** – Parameters obtained from SANS analysis, as illustrated in the main manuscript. For different salt concentration, $\phi_M$, the number of layers in MLVs, $N$, the outer perimeter radius, $R_{SANS}$, and the thickness of the water layer between the bilayers $t_w$.

| $\phi_M$ (mM) | $R_{SANS}$ (Å) | $\delta_{HH}$ (Å) | $t_w$ (Å) | $N$ |
|---|---|---|---|---|
| 0 | $539 \pm 8$ | $39 \pm 2$ | NA | 1 |
| 2 | $530 \pm 9$ | $40 \pm 2$ | NA | 1 |
| 4 | $516 \pm 11$ | $38 \pm 3$ | $20 \pm 2$ | 2 |
| 6 | $498 \pm 8$ | $37 \pm 2$ | $20 \pm 1$ | 2 |
| 8 | $549 \pm 10$ | $38 \pm 2$ | $22 \pm 1$ | 2 |
| 10 | $539 \pm 12$ | $40 \pm 2$ | $20 \pm 2$ | 2 |
| 20 | $532 \pm 11$ | $42 \pm 2$ | $18 \pm 2$ | 2 |
| 40 | $525 \pm 8$ | $38 \pm 2$ | $19 \pm 2$ | 2 |
| 150 | $504 \pm 8$ | $39 \pm 2$ | $15 \pm 2$ | 3 |
| 470 | $502 \pm 16$ | $38 \pm 2$ | $14 \pm 2$ | 3 |

**Table S2** – Parameters obtained SAXS analysis, as illustrated in Figure 4 of the main manuscript. For different salt concentration, $\phi_M$, the number of layers, $N$, average lamellar spacing, $d$, size of the head, $\delta_H$, the head-to-head bilayer thickness, $\delta_{HH}$, the thickness of the water layer between the bilayers $t_w$, and the Caille parameter, $\eta_{cp}$.

| $\phi_M$ (mM) | $N$ | $d$ (Å) | $\delta_H$ (Å) | $\delta_{HH}$ (Å) | $t_w$ (Å) | $\eta_{cp}$ |
|---|---|---|---|---|---|---|
| 0 | 1 | 63 ± 1 | 6.7 ± 0.5 | 41.37 ± 1.2 | NA | 0.10 ± 0.01 |
| 6 | 3 ± 1 | 64 ± 1 | 7.1 ± 0.3 | 40.78 ± 2.2 | 22.70 ± 1.2 | 0.10 ± 0.01 |
| 10 | 3 ± 1 | 62 ± 1 | 7.8 ± 0.8 | 42.66 ± 3.0 | 20.00 ± 2.0 | 0.11 ± 0.02 |
| 20 | 3 ± 1 | 60 ± 2 | 7.8 ± 0.8 | 38.80 ± 1.0 | 21.20 ± 0.6 | 0.11 ± 0.01 |
| 40 | 3 ± 1 | 59 ± 1 | 6.1 ± 0.4 | 38.06 ± 1.2 | 20.57 ± 0.2 | 0.10 ± 0.01 |
| 150 | 4 ± 1 | 56 ± 1 | 6.0 ± 0.5 | 39.3 ± 1.4 | 16.85 ± 0.4 | 0.18 ± 0.02 |
| 470 | 4 ± 1 | 54 ± 1 | 6.0 ± 0.5 | 39.47 ± 1.2 | 15.06 ± 0.2 | 0.20 ± 0.02 |